\documentclass[preprintnumbers,superscriptaddress,pra]{revtex4}
\usepackage{amssymb}
\usepackage{amsmath}
\usepackage{epsfig}
\usepackage{graphicx}
\usepackage{color}

\input{tcilatex}
\begin{document}

\title[Gravitational area contraction in Mercury orbit]{Gravitational
major-axis contraction of Mercury's elliptical orbit}
\author{Q. H. Liu}
\email{quanhuiliu@gmail.com}
\affiliation{School for Theoretical Physics, School of Physics and Electronics, Hunan
University, Changsha 410082, China}
\affiliation{Synergetic Innovation Center for Quantum Effects and Applications (SICQEA),
Hunan Normal University,Changsha 410081, China}
\author{Q. Li}
\affiliation{School for Theoretical Physics, School of Physics and Electronics, Hunan
University, Changsha 410082, China}
\author{T. G. Liu}
\affiliation{School for Theoretical Physics, School of Physics and Electronics, Hunan
University, Changsha 410082, China}
\author{X. Wang}
\affiliation{School for Theoretical Physics, School of Physics and Electronics, Hunan
University, Changsha 410082, China}
\date{\today }

\begin{abstract}
The local curvature of the space produced by the Sun causes not only the
perihelion precession of Mercury's elliptical orbit, but also the variations
of the whole orbit, in comparison with those predicted by the Newtonian
theory of gravitation. Calculations show that the gravitational major-axis
contraction of the Mercury's elliptical orbit is $1.3$ kilometers which can
be confirmed by the present astronomical distance measurement technology.
\end{abstract}

\maketitle

\section{Introduction}

Mercury is the planet closest to the Sun, the swiftest in the solar system 
\cite{ESO,nasa}. She revolves around the Sun in an elliptical orbit at a
mean distance of about $57.9$ million km; the orbital period is $0.24$ Earth
years \cite{ESO,nasa}. At perihelion the Mercury is $r_{p}\approx 46.0$
million km from the Sun, and at aphelion it is at a distance of $%
r_{a}\approx 69.8$ million km, and the orbital eccentricity $e$ which is
rather large ($e\approx 0.2056$) \cite{ESO,nasa} in comparison with that of
the Earth ($e\approx 0.0167$). It was a historical puzzling that there was
from the Newtonian theory of gravitation an anomalous precession of the
perihelion of Mercury, about $43$ seconds of arc per century, and in 1915
Einstein proposed to take such a procession as an observational evidence for
the test of his theory of general relativity \cite{weinberg}. Nowadays, it
is the standard approach that the gravitation must be understood within the
general relativity. A curious question on the gravitational influence on the
Mercury's elliptical orbit is in the following. For an elliptical orbit with
Sun at one focus, one can conveniently choose the focus as origin of polar
coordinates, and there are two coordinates, polar angular and radial ones,
usually denoted by ${\varphi }$ and $r$, respectively, to specify the
position of the planet. The gravitation must simultaneously have effects on
both the angular and radial motions. We can ascertain that the
gravitationally induced change of the radial motion is small, and had been
hard to be observed when the general relativity was establishing in 1915 and
some years later, and it may be testable today. The present paper is devoted
to study the gravitational major-axis contraction of Mercury's elliptical
orbit. In the present paper, we consider only the effects of the local
curvature of the space produced by the Sun on the motion of the Mercury,
without taking other gravitational perturbations coming from, e.g., other
planets and oblateness of the Sun etc., into account.

This paper is organized in the following. In section II, the gravitational
major-axis contraction of Mercury's elliptical orbit is defined as the
difference between two (proper) distances between the perihelion$\ $and the
aphelion of the Mercury's orbit, predicted by the Einstein's and Newtonian
theory of the gravitation, respectively. In section III, the new perihelion $%
r_{P}\ $and aphelion $r_{A}$ of the Mercury's orbit in the Einstein's theory
of gravitation are calculated.\ In final section IV, a brief conclusion is
given.

\section{Definitions of two lengths of the major-axis of the Mercury's
elliptical orbit in modern and classical theory of gravitation}

For an ellipse whose equation is, with $p\equiv a(1-e^{2})$ and $a$ standing
for the semi-major axis, 
\begin{equation}
r=\frac{p}{1+e\cos {\varphi }},  \label{ellipse}
\end{equation}%
the area $S=\pi a^{2}\sqrt{1-e^{2}}$, and the perimeter $C=4aE(e)$ with $E(e)
$ denoting the complete elliptic integral of the second kind. For a circle
of radius $a$, we have with $e=0$, $S=\pi a^{2}$ and $C=4aE(0)=$ $2\pi a$
with $E(0)=\pi /2$. 

In the Newtonian theory of gravitation for planets in the solar system, the
elliptical orbit satisfies following Binet's equation, \cite{apple}%
\begin{equation}
\frac{{{d}^{2}}u}{d{{\varphi }^{2}}}+u={{b}^{2},}  \label{BinetEq}
\end{equation}%
where $u=r_{gs}/r$ with $r_{gs}\equiv GM/c^{2}$ denoting the gravitational
radius of the Sun, and $GM$ being the product of gravitational constant and
the mass of the Sun and $c$ being the velocity of light speed, and $%
b=GM/(cL)=r_{gs}c/L$ with $L$ being the dimensionless specific angular
momentum. For the Sun, $r_{gs}\sim 1.47$ km, and for Mercury, mean value of
radius is $\bar{r}\sim 57.9$ million km, we have ${{b}^{2}\sim }\bar{u}%
=r_{gs}/\bar{r}\approx 0.25\ast 10^{-7}$. The solution to equation (\ref%
{BinetEq}) is,%
\begin{equation}
u=u_{0}\equiv {{b}^{2}(1+e\cos \varphi ).}  \label{kepler}
\end{equation}%
This result is purely a geometric one, given by the Newtonian theory of
gravitation. The perihelion $r_{p}\ $and aphelion $r_{a}$ are, respectively, 
\begin{equation}
r_{p}=\frac{1}{{{b}^{2}(1+e)}}\approx 46.0\text{ million km, and }r_{a}=%
\frac{1}{{{b}^{2}(1-e)}}\approx 69.8\text{ million km.}
\end{equation}
However, the astronomical observations indicate that the orbit is not
exactly closed as suggested by (\ref{kepler}), and the Einstein's theory of
gravitation must be utilized to understand the anomalous behavior of the
Mercury's orbit.

The Einstein's theory of gravitation presents that the external
gravitational field produced by the Sun has spherical symmetry and is
usually expressed by the Schwarzschild metric, \cite{apple,weinberg,landau} 
\begin{equation}
ds^{2}=-\left( 1-\frac{2r_{gs}}{r}\right) dt^{2}+\frac{1}{\left( 1-\frac{%
2r_{gs}}{r}\right) }dr^{2}+r^{2}\left( d\theta ^{2}+\sin ^{2}\theta d{%
\varphi }^{2}\right) {,}  \label{metric}
\end{equation}%
where the parameter $s$ denotes the proper length, the parameter $t$ is the
coordinate time and $(r,\theta ,{\varphi )}$ are the conventional spherical
coordinates. When the Mercury moves in the gravitational field of the Sun,
as usual, we choose $\theta =\pi /2$ as the orbit plane. We can then use the
Einstein's theory of gravitation to get the new perihelion $r_{P}\ $and the
new aphelion $r_{A}$ of the Mercury's orbit. Once these two quantities are
known, the proper distance between the perihelion and the aphelion, or the
new length of the major-axis $D$, is approximately given by,%
\begin{equation}
D\approx \int_{R_{\odot }}^{r_{P}}\frac{1}{\sqrt{1-\frac{2r_{gs}}{r}}}%
dr+\int_{R_{\odot }}^{r_{A}}\frac{1}{\sqrt{1-\frac{2r_{gs}}{r}}}dr+2R_{\odot
}\approx r_{P}+r_{A}+r_{gs}\ln \frac{r_{P}r_{A}}{R_{\odot }^{2}},
\label{properdistance}
\end{equation}%
where $R_{\odot }\approx 69.6\ast 10^{-2}$ million km is the mean radius of
the Sun. Evidently, the new major-axis $D$ in the Einstein's theory of
gravitation comes from two parts, 
\begin{equation}
r_{P}+r_{A}\text{ and }r_{gs}\ln \frac{r_{P}r_{A}}{R_{\odot }^{2}}.
\label{2p}
\end{equation}%
The first part $r_{P}+r_{A}$ will be computed in the following section, and
turns out to be less than $r_{p}+r_{a}$. The second part proves to be
approximately, with $r_{p}\approx 46.0$ and $r_{a}\approx 69.8$ million km\
in the Newtonian theory of gravitation, 
\begin{equation}
r_{gs}\ln \frac{r_{P}r_{A}}{R_{\odot }^{2}}\approx r_{gs}\ln \frac{r_{p}r_{a}%
}{R_{\odot }^{2}}\approx 12.9\text{ km.}  \label{sundiameter}
\end{equation}%
This mainly results from the curvature of the space induced by the
gravitational field of the Sun.

Remember that the Newtonian theory predicts an exactly closed elliptic orbit
for the Mercury, but the Einstein's theory correctly gives the\ precession
the perihelion. Though the gravitational length contraction is hard to
define in general, the gravitational major-axis contraction of Mercury's
elliptical orbit can be well-defined in the similar manner. On one hand, we
have the length of the major-axis predicted by the Newtonian theory of
gravitation, which is $r_{p}+r_{a}$. On the other, we have that length from
the Einstein's theory, which is $D$ (\ref{properdistance}). Their difference
is clearly given by,%
\begin{equation}
D-(r_{p}+r_{a})=r_{P}+r_{A}-(r_{p}+r_{a})+r_{gs}\ln \frac{r_{P}r_{A}}{%
R_{\odot }^{2}},
\end{equation}
which will be shortly shown to be $-1.3$ kilometers, which is referred as to
the gravitational major-axis contraction of Mercury's elliptical orbit. In
following section, we will calculate $r_{P}$ and $r_{A}$.

\section{New perihelion $r_{P}\ $and aphelion $r_{A}$ of the Mercury's orbit
in the Einstein's theory of gravitation}

Once the gravitation is static and weak, the Einstein's theory of general
relativity predicts that Mercury's orbit satisfies following nonlinear
equation, an extension of the Binet's equation,%
\begin{equation}
\frac{{{d}^{2}}u}{d{{\varphi }^{2}}}+u={{b}^{2}}+3u^{2}.  \label{ExBinetEq}
\end{equation}%
In this equation, the quantity $3u^{2}\sim 0.19\ast 10^{-14}$ is a
perturbation in comparison with {the first term on the right handed side }${{%
b}^{2}\sim }u\approx 0.25\ast 10^{-7}$. This equation is still a geometrical
one expressing the relation between angular variable ${\varphi }$ and the
inverse of the dimensionless radial variable $u$. Without the presence of
the perturbation, the Binet's equation (\ref{BinetEq}) gives a perfect
ellipse, and the effective frequency is unity, as shown in Eq. (\ref{kepler}%
). With the perturbation, one can no longer expect that the frequency
remains unity. To solve Eq. (\ref{ExBinetEq}), we utilize the standard
perturbation method \cite{PT} to get an approximate solution. For sake of
convenience, we introduce a real parameter $\varepsilon $ into equation (\ref%
{ExBinetEq}) which then becomes, 
\begin{equation}
\frac{{{d}^{2}}u}{d{{\varphi }^{2}}}+u={{b}^{2}}+3\varepsilon u^{2},
\label{perturb1}
\end{equation}%
which can also be rewritten into following form with a new parameter $\xi
\equiv \omega \varphi $ ($\xi \in \lbrack 0,\infty )$), where $\omega $
differs from unity due to the possibly small frequency renormalization as a
consequence of removal of the secular term as shown in (\ref{ansatz2}), 
\begin{equation}
\omega ^{2}\frac{{{d}^{2}}u}{d\xi {^{2}}}+u={{b}^{2}}+3\varepsilon u^{2}.
\label{peturb2}
\end{equation}%
In deriving Eq. (\ref{ExBinetEq}), there is an effective radial confining
potential \cite{apple,weinberg,landau}, from which we know that the particle
is bounded between the aphelion and perihelion, and every bound state
possesses a fixed frequency. Thus $\omega $ must also be a constant. The
perturbation solution to the equation (\ref{peturb2}) can be expanded in
powers of $\varepsilon $, \cite{PT} 
\begin{subequations}
\begin{eqnarray}
u &=&{{u}_{0}}(\xi )+\varepsilon {{u}_{1}}(\xi )+{{\varepsilon }^{2}}{{u}_{2}%
}(\xi )+\cdots   \label{ansatz1} \\
\omega  &=&1+\varepsilon {{\omega }_{1}}+{{\varepsilon }^{2}}{{\omega }_{2}}%
+\cdots   \label{ansatz2}
\end{eqnarray}%
where the subscripts $0,1,2,...$in expansions of $u$ and $\omega $ denote
the orders of approximation. When all required calculations are complete, we
set $\varepsilon =1$, so above expansions are in fact in powers of $u\sim {{b%
}^{2}}$, as it should be.

The zeroth order solution of the elliptical orbit is already given by the
Newtonian theory of gravitation, i.e., (\ref{kepler}). The first order
solution\ ${{u}_{1}}(\varphi )$ satisfies following equation, 
\end{subequations}
\begin{equation}
\frac{{{d}^{2}}{{u}_{1}}}{d\xi {^{2}}}+{{u}_{1}}=3u_{0}^{2}-2{{\omega }_{1}}{%
{{u}^{\prime \prime }}_{0}}=3u_{0}^{2}-2{{\omega }_{1}}({{b}^{2}}-{{u}_{0}}).
\label{1storder}
\end{equation}%
The initial conditions are, \cite{PT} 
\begin{equation}
{{u}_{1}(0)=0,}\text{ }\frac{{d}{{u}_{1}(0)}}{d\xi }=0.  \label{initial}
\end{equation}%
The solution to equation (\ref{1storder}) is, 
\begin{equation}
{{u}_{1}}=2{{b}^{4}}(3+2{{e}^{2}}+{{e}^{2}}\cos \xi ){{\sin }^{2}}\left( \xi
/2\right) +{{b}^{2}}e({{\omega }_{1}}+3{{b}^{2}})\xi \sin \xi {.}
\label{secular}
\end{equation}%
The second term is secular for it contains $\xi \sin \xi $ which can be
infinitely large as $\xi \rightarrow \infty $ and must be eliminated by
setting, 
\begin{equation}
{{\omega }_{1}}=-3{{b}^{2}.}  \label{normalization}
\end{equation}%
which accounts for the procession of the perihelion of Mercury. \cite%
{apple,weinberg,landau} To see it explicitly, we maximize $u\simeq {{u}_{0}}%
(\xi )+{{u}_{1}}(\xi )={{b}^{2}(1+e\cos \xi )+}2{{b}^{4}}(3+2{{e}^{2}}+{{e}%
^{2}}\cos \xi ){{\sin }^{2}}\left( \xi /2\right) $ which is to be ${{b}%
^{2}(1+e)}$ that defines the perihelion of Mercury, when the parameter ${%
\theta }$ takes following values, 
\begin{equation}
\xi \equiv \omega \varphi =2\pi n,(n=0,1,2),  \label{parameter}
\end{equation}%
which can be rewritten in terms of the polar angle $\varphi $, 
\begin{equation}
\left( 1-3{{b}^{2}}\right) \varphi \approx 2\pi n.  \label{period}
\end{equation}%
This reproduces the well-known result for the angular difference between two
successive points of closest approach of Mercury to the Sun, \cite%
{apple,weinberg,landau} 
\begin{equation}
\Delta \varphi \approx 6\pi {{b}^{2}}.  \label{procession}
\end{equation}

The first order correction of $\varepsilon $ to the inverse of\ the
dimensionless radial position $u=r_{gs}/r$ is positive from (\ref{secular}),%
\begin{equation}
{{u}_{1}}=2{{b}^{4}}(3+2{{e}^{2}}+{{e}^{2}}\cos \xi ){{\sin }^{2}}\left( \xi
/2\right) \geq 0.  \label{1stcorrection}
\end{equation}%
Combination of this result and the unperturbed one (\ref{kepler}) gives, 
\begin{equation}
u\approx {{b}^{2}}(1+e\cos \xi )+2{{b}^{4}}(3+2{{e}^{2}}+{{e}^{2}}\cos \xi ){%
{\sin }^{2}}\left( \xi /2\right)   \label{appsol}
\end{equation}%
The radial position is then, 
\begin{subequations}
\begin{eqnarray}
r &\approx &r_{gs}\left( {{b}^{2}}(1+e\cos \xi )+2{{b}^{4}}(3+2{{e}^{2}}+{{e}%
^{2}}\cos \xi ){{\sin }^{2}}\left( \xi /2\right) \right) ^{-1}
\label{radial1} \\
&\approx &\frac{r_{gs}}{{{b}^{2}}(1+e\cos \xi )}\left( 1-\frac{2{{b}^{2}}(3+2%
{{e}^{2}}+{{e}^{2}}\cos \xi ){{\sin }^{2}}\left( \xi /2\right) }{1+e\cos \xi 
}\right)   \label{radial2} \\
&\approx &r_{gs}\left( \frac{1}{{{b}^{2}}(1+e\cos \xi )}-\frac{2(3+2{{e}^{2}}%
+{{e}^{2}}\cos \xi ){{\sin }^{2}}\left( \xi /2\right) }{\left( 1+e\cos \xi
\right) ^{2}}\right) .  \label{radial3}
\end{eqnarray}%
This equation gives both the coordinate of the new perihelion $r_{P}$, 
\end{subequations}
\begin{equation}
r_{P}=r_{\min }=\left. r\right\vert _{\xi =0}\approx \frac{r_{gs}}{{{b}^{2}}%
(1+e\cos \xi )}=r_{p},  \label{rp}
\end{equation}%
and the new aphelion $r_{A}$, 
\begin{equation}
r_{A}=r_{\max }=\left. r\right\vert _{\xi =\pi }\approx \frac{r_{gs}}{{{b}%
^{2}}(1-e)}-r_{gs}\frac{2(3+{{e}^{2}})}{\left( 1-e\right) ^{2}}=r_{a}-14.2%
\text{ km.}  \label{ra}
\end{equation}%
Combining results (\ref{rp}) and (\ref{ra}), we find that the distance
between the perihelion and the aphelion, i.e., the length of the major-axis
of the ellipse, is shortened, and the difference is, 
\begin{equation}
\Delta r\equiv r_{P}+r_{A}-(r_{p}+r_{a})=-r_{gs}\frac{2(3+{{e}^{2}})}{\left(
1-e\right) ^{2}}\approx -14.2\text{ km.}  \label{major}
\end{equation}%
In consequence, we see that result (\ref{sundiameter}) holds true.

Now we are ready to give the gravitational major-axis contraction of the
Mercury's elliptical orbit in unit of kilometer,%
\begin{equation}
D-(r_{p}+r_{a})\approx -14.2+12.9=-1.3.
\end{equation}%
In a little more detail, in the Newtonian theory of gravitation, we have
major-axis $(r_{p}+r_{a})$ with the Sun as a point of mass; however, in the
Einstein's theory, we have a little bit shorter one $D$, which can be
observed with present astronomical distance measurement technology. In the
website of NASA, \cite{nasa} we see that the distance of the Mercury from
the Sun is reported in every second with accuracy of $one$ kilometer.

\section{Conclusion}

To sum up, since the gravitation between Mercury and the Sun must have
influences on both radial and angular positions, we compute the
gravitational major-axis contraction of Mercury's elliptical orbit. The
result is $-1.3$ km which can be confirmed in the observation of the
Mercury's orbit, 

\begin{acknowledgments}
This work is financially supported by National Natural Science Foundation of
China under Grant No. 11675051. We are indebted to an anonymous reviewer for
pointing us the proper distance (\ref{properdistance}), which greatly
contributed to improving the final version of the paper.
\end{acknowledgments}

\end{document}